\documentclass[runningheads]{llncs}
\usepackage[T1]{fontenc}
\usepackage{subcaption}
\usepackage{graphicx}
\usepackage{booktabs}
\usepackage{amsmath}
\usepackage{comment}
\usepackage{multirow}
\usepackage{algpseudocode}
\usepackage{algorithm}
\usepackage{tikz}
\usepackage{amsfonts}
\DeclareMathOperator*{\argmin}{arg\,min}
\begin{document}
\let\origtau\tau
\renewcommand{\tau}{\scalebox{1.6}{$\origtau$}}

\title{Enhancing Multi-Exposure High Dynamic Range Imaging with Overlapped Codebook for Improved Representation Learning}
\titlerunning{OLC-HDR}
%
\author{Keuntek Lee\inst{1} \and
Jaehyun Park\inst{2} \and
Nam Ik Cho\inst{1,2}}
\authorrunning{Lee et al.}
%
\institute{
Department of ECE, INMC, Seoul National University, Seoul, Korea\\
\email{\{leekt000,nicho\}@snu.ac.kr} \and
IPAI, INMC, Seoul National University, Seoul, Korea\\
\email{jaep970805@gmail.com}}
\maketitle              
\begin{abstract}
High dynamic range (HDR) imaging technique aims to create realistic HDR images from low dynamic range (LDR) inputs. Specifically, Multi-exposure HDR imaging uses multiple LDR frames taken from the same scene to improve reconstruction performance. However, there are often discrepancies in motion among the frames, and different exposure settings for each capture can lead to saturated regions. In this work, we first propose an Overlapped codebook (OLC) scheme, which can improve the capability of the VQGAN framework for learning implicit HDR representations by modeling the common exposure bracket process in the shared codebook structure. Further, we develop a new HDR network that utilizes HDR representations obtained from a pre-trained VQ network and OLC. This allows us to compensate for saturated regions and enhance overall visual quality. We have tested our approach extensively on various datasets and have demonstrated that it outperforms previous methods both qualitatively and quantitatively.

\keywords{Exposure fusion  \and HDR imaging \and Vector quantization.}
\end{abstract}
\section{Introduction}
\label{sec:intro}
The task of multi-exposure high dynamic range (HDR) imaging is to create a high-quality HDR image from multiple low dynamic range (LDR) images that were taken with different exposure settings. This approach is superior to single-image HDR imaging, which lacks information and produces lower-quality results. By utilizing more information from multiple frames when LDR frames are perfectly still, multi-exposure HDR imaging can produce finer HDR results. However, LDR frames taken by exposure bracketing have motion differences from each other, and each LDR image has over- or under-exposed regions, which can lead to undesirable artifacts such as ghosting and washed-out areas in the final HDR image. To deal with these issues, earlier works~\cite{kalantari2017deep,zimmer2011freehand,wu2018deep} used pre-processing steps to align the LDR frames before merging them, by using optical flow or homography transformation. However, such explicit alignment methods can have estimation errors, bringing misaligned frames to the following merging stage. 

Recently, convolutional neural networks (CNNs) have achieved notable successes in various computer vision areas, including HDR imaging. Kalantari \textit{et al.}~\cite{kalantari2017deep} first proposed a CNN-based merging network for multi-exposure HDR imaging. Yan \textit{et al.}~\cite{yan2019attention} proposed an attention-based network that implicitly aligns non-reference frames at the feature level. More recently, Niu \textit{et al.}~\cite{niu2021hdr} proposed an HDR method based on the generative adversarial network (GAN)~\cite{goodfellow2020generative}, and Liu \textit{et al.}~\cite{liu2022ghost} presented an algorithm based on the Vision Transformer (ViT)~\cite{dosovitskiy2020image}. Although CNN-based methods generally outperform traditional methods in HDR reconstruction, they still struggle with saturated regions and missing details on severely under-/over-exposed LDR frames.

\begin{figure*}[t]
\centering
\centering
\begin{subfigure}[b]{0.31\textwidth}
   \includegraphics[width=\textwidth]{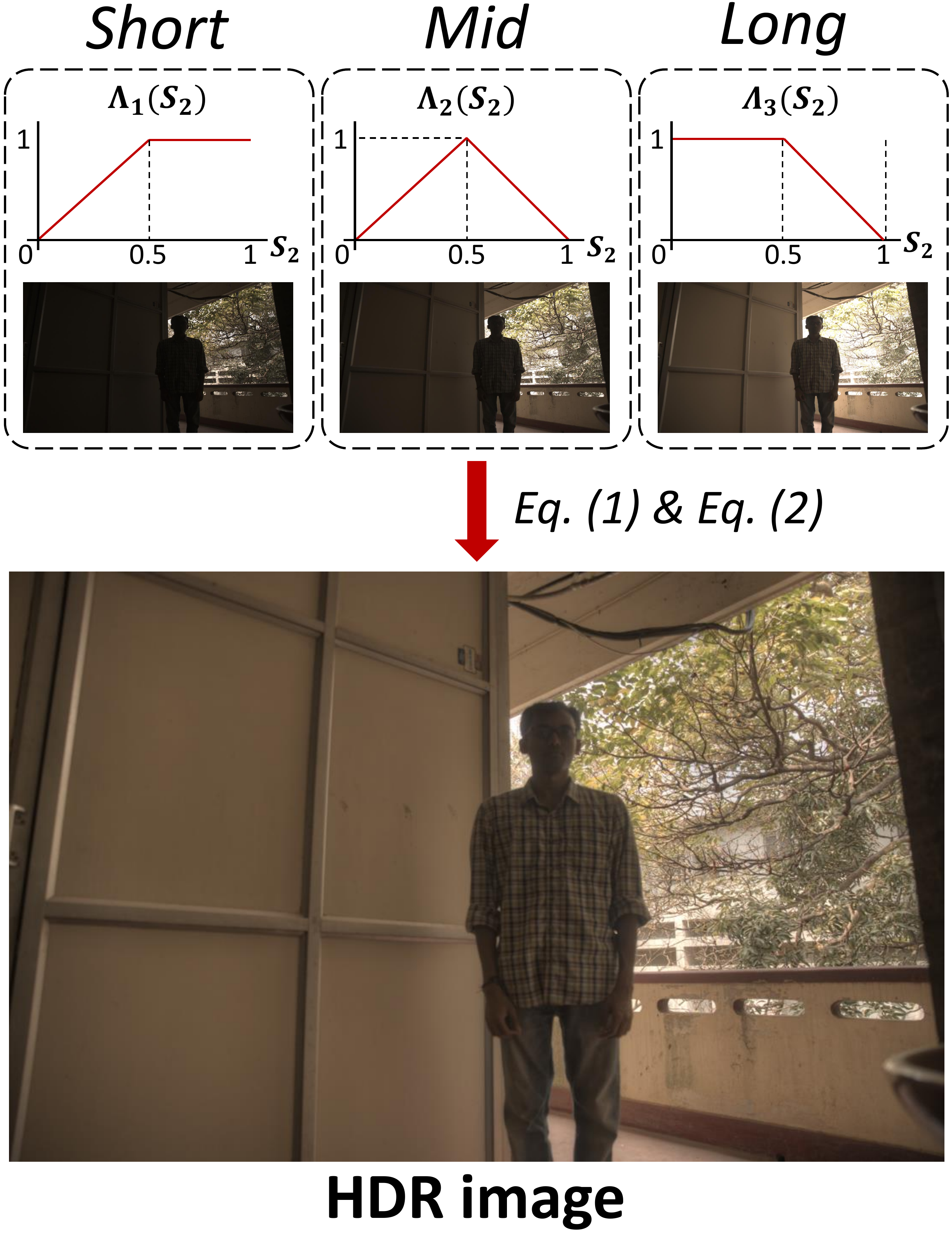}
   \caption{Exposure bracketing}
\end{subfigure}\hfill
\begin{subfigure}[b]{0.65\textwidth}
   \includegraphics[width=\textwidth]{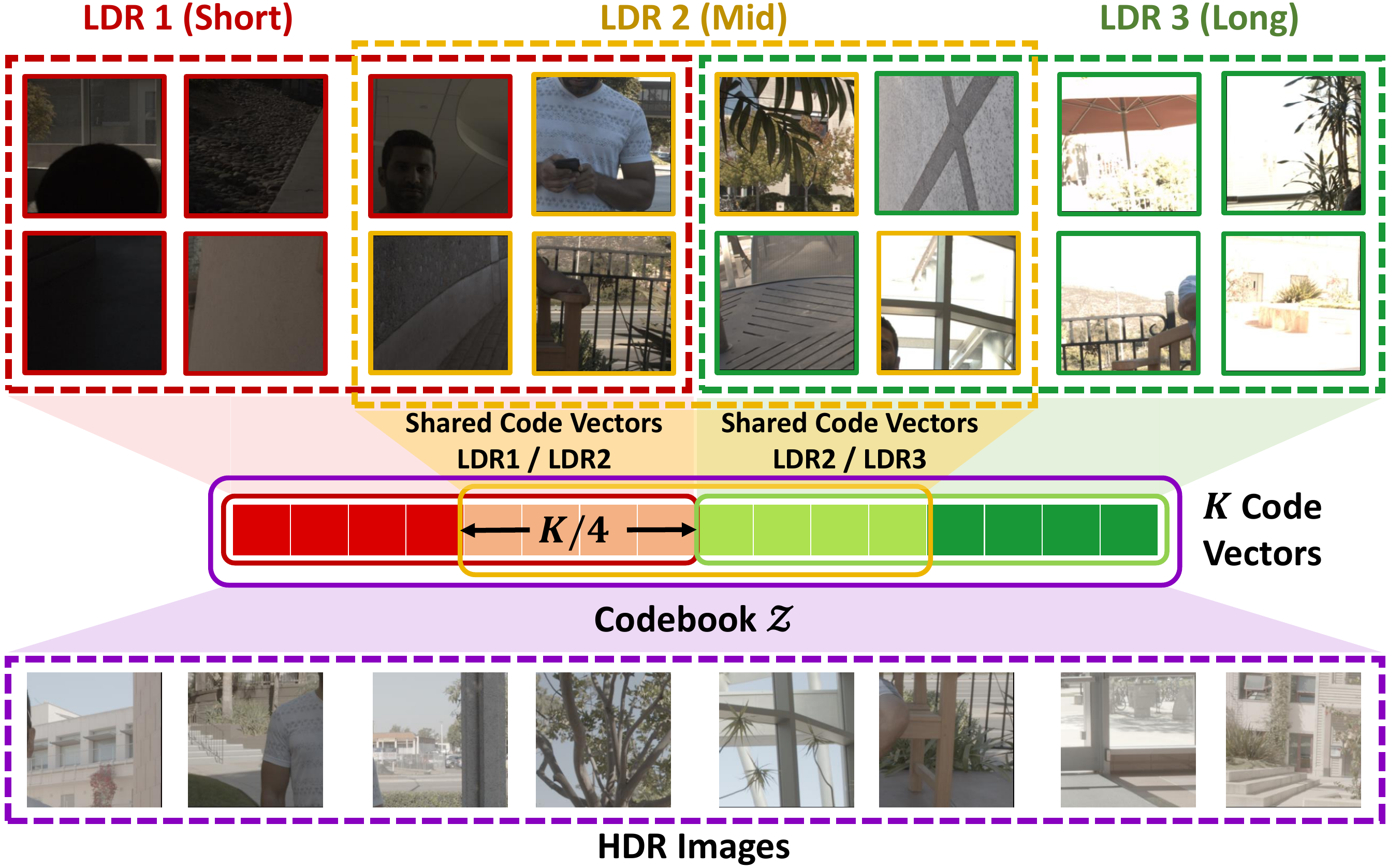}
   \caption{Proposed Overlapped Codebook (OLC) scheme}
\end{subfigure}
\caption{Illustration of (a) conventional exposure bracketing process with triangle function ($\Lambda_{1},\Lambda_{2},\Lambda_{3}$) and (b) proposed Overlapped Codebook (OLC) scheme for multi-exposure HDR imaging. The proposed OLC scheme is able to represent HDR images with a combination of LDR representations by aligning the exposure bracket process with its codebook structure.} 
\label{fig:OLC}
\vspace{-1.5em}
\end{figure*}

In this work, we introduce a novel HDR reconstruction network with a dual-decoder structure that leverages learned HDR representations to restore fine details and compensate for saturated regions. Our approach employs a vector quantization (VQ) mechanism for learning HDR image representations, specifically proposing the Overlapped Codebook (OLC) scheme that models the exposure bracket fusing process (Fig.\ref{fig:OLC}(a)). The proposed OLC learns LDR frame representations within specific codebook segments based on exposure bias (short, mid, long) while utilizing the full codebook for HDR priors, enhancing the learning of implicit HDR representations (Fig.\ref{fig:OLC}(b)). This scheme allows the proposed OLC to represent HDR information by combining LDR representations, similar to the traditional exposure bracket process. The HDR network integrates latent features from the pre-trained VQ decoder and frame context into the fidelity decoder a residual fusing modules, improving HDR image quality. To address frame misalignment, we introduce a parallel alignment module and a dynamic frame merging module to combine LDR frame context with valid regional features. These components collectively enhance the HDR reconstruction process. Experimental results demonstrate that our method outperforms previous methods across various datasets and metrics.

Our contributions can be summarized as follows:

\begin{itemize}

    \item We introduce an Overlapped Codebook (OLC) scheme for implicitly capturing HDR representations via the VQGAN framework. The OLC aligns with the common exposure bracketing process, achieving improved representation learning ability for multi-exposure HDR imaging.

    \item We present a dual-decoder HDR network, integrating learned HDR representations from a pre-trained VQ decoder and OLC into the fidelity decoder for high-quality HDR image generation. Additionally, we introduce a parallel alignment module and a frame-selective merging module to address misalignment and incorporate frame context effectively.

    \item Extensive experiments demonstrate that our HDR network with learned representation in pre-trained OLC achieves superior performance on various datasets and metrics. 
    

    
\end{itemize}

\section{Related Works}
\label{sec:relatedworks}

\subsection{Multi-Exposure HDR imaging}
Multi-exposure HDR imaging generally produces higher-quality results compared to single-image HDR imaging. This is because it can leverage more information from multiple LDR frames. However, taking multiple LDR images can cause hand or object motions, and some LDR images may have under-/over-exposed regions due to scene conditions and exposure biases. Therefore, aligning LDR frames and compensating for saturated areas are the primary concerns in multi-exposure HDR imaging schemes.

Earlier methods proposed a pixel rejection approach for multi-exposure HDR imaging, assuming the images are globally registered. For instance, Grosch~\cite{grosch2006fast} uses the color difference of input images as an error map. Jacobs \textit{et al.}~\cite{jacobs2008automatic} measure weighed variance for detecting ghost regions. The registration-based methods were also proposed, which search for similar regions. Kang \textit{et al.}~\cite{kang2003high} utilize exposure bias information to transform LDR images to the luminance domain and apply optical flow for finding corresponding pixels from non-reference LDR frames. Sen \textit{et al.}~\cite{sen2012robust} introduced a patch-based energy minimization method for jointly optimizing input alignment and HDR image reconstruction.

Recently, CNN-based methods have shown superior performance in various image restoration areas, including HDR imaging. Kalantari \textit{et al.} \cite{kalantari2017deep} first proposed a CNN-based method for multi-exposure HDR imaging. They adopted optical-flow estimation for aligning LDR frames in the pre-processing stage, then merged LDR images at the feature level. Wu \textit{et al.}~\cite{wu2018deep} aligned the background through the homography transformation and applied a network with skip-connection for merging. Yan \textit{et al.}~\cite{yan2019attention} proposed a network with a spatial attention module for aligning LDR frames implicitly in the feature domain. Non-local~\cite{wang2018non} method was also proposed by Yan \textit{et al}.~\cite{yan2020deep}, which constructs a non-local module and triple-pass residual module in the network bottleneck. More recently, Niu \textit{et al.}~\cite{niu2021hdr} proposed a GAN-based network for producing a more realistic result, which consists of a generator with reference-based residual merging block. Liu~\cite{liu2021adnet} employed a pyramid cascading deformable (PCD) module~\cite{wang2019edvr} to align frame features. Vision Transformer (ViT)~\cite {dosovitskiy2020image} has also achieved impressive performance in image restoration areas \cite{liu2021swin,zamir2022restormer}, and thus applied to HDR imaging. Liu \textit{et al.}~\cite{liu2022ghost}, Chen \textit{et al.}~\cite{chen2023improving} and Yan \textit{et al.}~\cite{yan2023unified} introduce Transformer-based models for capturing the complex relationship between LDR frames. Further, Song \textit{et al.}~\cite{song2022selective} proposed a Transformer network with a ghost region detector to make the network focus on valid regions. Tel \textit{et al.}~\cite{tel2023alignment} introduced an inter-/intra-frame merging Transformer network with a cross-attention mechanism for utilizing spatial and semantic information.

\subsection{Vector Quantization}
VQ-VAE~\cite{van2017neural} was the first to introduce a VQ mechanism to neural networks, which learns discrete code vectors for encoding images. Recently, Esser \textit{et al.}~\cite{esser2021taming} proposed VQGAN for achieving high-quality generated images, which trains the codebook over Transformer architecture and adversarial objectives. The VQ mechanism has also been widely adopted in image restoration areas. Guo \textit{et al.}~\cite{guo2022lar} proposed a super-resolution method with a texture codebook and local autoregressive model for producing finer details. Chen \textit{et al.}~\cite{chen2022real} introduced a super-resolution network with the pre-trained codebook to leverage learned high-resolution priors. Gu \textit{et al.}~\cite{gu2022vqfr} proposed a face restoration network that takes advantage of the high-quality feature in the VQ codebook to produce images with realistic face details.

\section{Proposed Methods}
\label{sec:methods}

Given a set of LDR frames with different exposure biases, our target is to compose a single HDR image by the best use of LDR frames' information. Specifically, we propose a 2-step method for multi-exposure HDR imaging which can be summarized as follows: 
\begin{itemize}
    \item \textbf{Step 1, Learning implicit HDR representations with the Overlapped Codebook (OLC). }
    \item \textbf{Step 2, HDR reconstruction with the pre-trained OLC and VQ decoder}.
\end{itemize}
The details of each step are described in the following subsections.



\subsection{Learning HDR representation with the OLC}
In this section, we present an OLC, a method that enhances the learning process for capturing HDR representation by aligning with the HDR image generation process. The traditional method for creating ground-truth images in multi-exposure HDR imaging tasks involves merging captured bracketed exposure images~\cite{debevec2023recovering,kalantari2017deep}. For instance, Kalantari \textit{et al}.~\cite{kalantari2017deep} employed a triangular weighting function to blend differently exposed static LDR images ($S_{1},S_{2},S_{3}$) as:
\begin{equation}
H = \frac{\sum_{i}\alpha_{i}(S_{i}^{\gamma}/t_{i})}{\sum_{i}\alpha_{i}}, i=1,2,3,
\label{triangle}
\end{equation}
where $H$ is the generated HDR image, $\gamma$ is a parameter for the gamma-correction function. The $\alpha_{i}$ is the weights for each LDR frame, which can be defined:
\begin{equation}
    \alpha_{1} = 1-\Lambda_{1}(S_{2}),\alpha_{2} = \Lambda_{2}(S_{2}),\alpha_{3} = 1-\Lambda_{3}(S_{2}),
\end{equation}
where $\Lambda_{i}(\cdot)$ is the triangle function described in Fig.~\ref{triangle}(a).
To reflect the above-stated weight blending process in multi-exposed LDR fusing, we propose the OLC method that concurrently learns LDR and HDR representations, forming HDR information through a combination of LDR representations. As illustrated in Fig. ~\ref{fig:OLC}(b), within the OLC framework, each LDR frame is linked to a specific codebook segment based on its exposure bias (short, mid, long) and shares codebook elements with other LDR frames. In contrast, the HDR image is represented using the entire codebook. This distinctive approach employed by OLC improves the capability to represent HDR images within VQ mechanisms.

\begin{figure*}[t]
\centering
\centering
   \includegraphics[width=12cm]{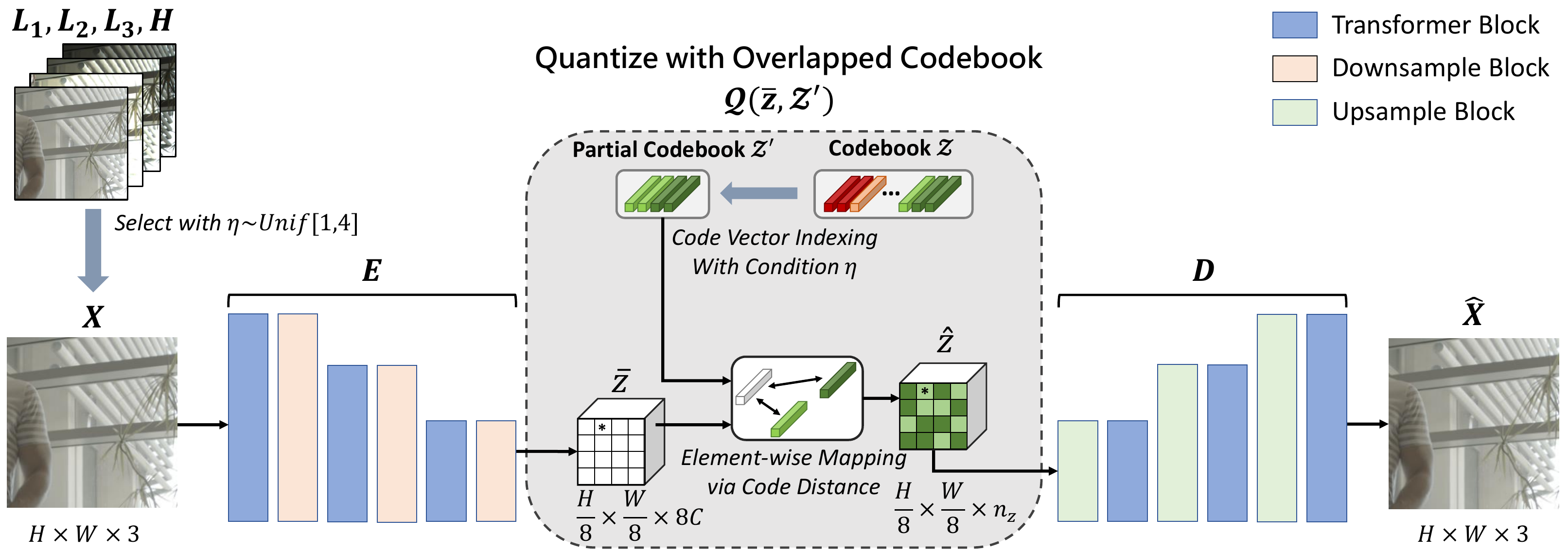}

\vspace{-0.5em}

\caption{Illustration of proposed Overlapped Codebook (OLC) scheme with VQGAN framework. In every iteration, we sample $\eta\sim Unif[1,4]$ for randomly selecting input image $X$ and indexing the corresponding codebook segment $\mathcal{Z}^{\prime}$.} 
\label{vqgan}
\vspace{-1.0em}
\end{figure*}

As illustrated in Fig. ~\ref{vqgan}, we employ the VQGAN framework~\cite{esser2021taming}, which consists of encoder $E$, decoder $D$, and the overlapped codebook $\mathcal{Z}=\{z_{k}\}_{k=1}^{K}\in\mathbb{R}^{K\times n_{z}}$, where $K$ is the codebook size and $n_{z}$ is the code vector dimension. Given an input image $X\in \mathbb{R}^{H\times W\times 3}$, the encoder produces feature $\bar{z}=E(X)\in \mathbb{R}^{h\times w\times n_{z}}$. Note that input image $X$ can be each frame of LDR images $L_i$, $i=1,2,3$ or HDR image $H$. We randomly select an input from those images with the uniformly sampled parameter $\eta\sim Unif[1,4]$, which can be defined as:
\begin{equation}
    \begin{aligned}
        X=\begin{cases}
        L_{\eta}^{\gamma} / t_{\eta}, &\eta\in\{1,2,3\}\\
        H, &\eta == 4
        \end{cases}\\
    \end{aligned}
\end{equation}
where $\gamma=2.2$ is the parameter of the function and $t_{\eta}$ is the exposure bias of the corresponding input LDR image. Note that we use a gamma-correction function on LDR inputs, which maps LDR images into the HDR domain to alleviate the discrepancy between LDR and HDR images.
Then, the vector-quantized feature $\hat{z}$ is obtained by finding the nearest neighbors of each feature element in the codebook $\mathcal{Z}$. Different from the common codebook in the VQ scheme, the proposed OLC uses a specific part of codebook $\mathcal{Z}$ following the type of input image $X$. For instance, when input image $X$ is one of LDR image frame $L_{i}$, partial codebook $\mathcal{Z}^{i}\in\mathbb{R}^{(K/2) \times{n_{z}}}$ can be defined as:
\begin{equation}
\mathcal{Z}^{i}=\{z_{i\times \alpha + 1},z_{i\times \alpha+2},...,z_{(i+1)\times \alpha}\}, i \in \{1,2,3\},
\end{equation}
where $\alpha=\frac{K}{4}$ is the offset parameter, and $i$ is the index of the LDR frame. When the input image $X$ is an HDR image $H$, all $K$ code vectors are used ($\mathcal{Z}$). Note that the codebook $\mathcal{Z}^{i}$ for each LDR frame shares $\frac{K}{4}$ of code vectors. For instance, in the case of partial codebook $\mathcal{Z}^{1}$, $\mathcal{Z}^{2}$ for $L_{1}, L_{2}$, they share code vectors $\{z_{\alpha+1 },z_{\alpha+2},...,z_{2\times\alpha}\}\in\mathbb{R}^{\alpha\times n_{z}}$. 
The VQ process for encoded feature $\bar{z}=E(X)$ can be formulated as:
\begin{equation}
\begin{aligned}
    \hat{z}_{j}=\mathcal{Q}(\bar{z}_{j},\mathcal{Z}^{\prime})=\argmin_{z_{k}\in\mathcal{Z}^{\prime}}{\|\bar{z}_{j}-z_{k}\|},\eta\in \{1,2,3,4\},\\
    \text{where } \mathcal{Z}^{\prime}=\begin{cases}
    \mathcal{Z}^{\eta},& \eta\in\{1,2,3\}\\
    \mathcal{Z},              &\eta==4
\end{cases}
\end{aligned}
\end{equation}
where $\mathcal{Q}(\cdot)$ is a quantization function conditioned by the partial codebook $\mathcal{Z}^{\prime}$, 
$\hat{z}\in\mathbb{R}^{h\times w \times n_{z}}$ is a quantized feature, and $j\in\{1,2,...,h\times w\}$. Then, the decoder $D$ reconstructs the result $\hat{X} \approx X$, which can be formulated as:
\begin{equation}
\hat{X}=D(\mathcal{Q}(E(X),\mathcal{Z}^{\prime}))\in\mathbb{R}^{H\times W\times 3}.
\end{equation}
Since the quantization function $\mathcal{Q}(\cdot)$ is non-differentiable, we follow previous works \cite{van2017neural,esser2021taming} for backpropagation, which simply copies the gradients from the decoder $D$ to the encoder $E$. Thus, the codebook, encoder, and decoder can be optimized with loss function $\mathcal{L}_{vq}$, $\mathcal{L}_{rec}$, and $\mathcal{L}_{per}$, which can be defined as:
\begin{equation}
\mathcal{L}_{vq}=\|\text{sg}[E(X)]-\hat{z}\|_{2}^{2}+\beta \|\text{sg}[\hat{z}]-E(X)\|_{2}^{2},
\label{vq_loss}
\end{equation}
where $\beta=0.25$ is the commitment weight and $\text{sg}[\cdot]$ is the stop-gradient operation. It is worth noting that our partial codebook $\mathcal{Z}^{\prime}$ uses a specific part of the codebook $\mathcal{Z}$ by indexing code vectors. Thus, updating $\mathcal{Z}^{\prime}$ with Eq. \ref{vq_loss} is the same as updating corresponding code vectors in the master codebook $\mathcal{Z}$. The reconstruction loss and perceptual loss are defined as follows:
\begin{equation}
\mathcal{L}_{rec}=\|\tau(X)-\tau(\hat{X})\|_{1}, \mathcal{L}_{per}=\|\phi(\tau(X))-\phi(\tau(\hat{X}))\|_{1},
\end{equation}
where $\tau(\cdot)$ is a $\mu$-law tone-mapping function, and $\phi(\cdot)$ is the pre-trained VGG-16~\cite{simonyan2014very} network. Note that we follow \cite{kalantari2017deep,yan2019attention,niu2021hdr} to train networks more effectively, which apply the tone-mapping function $\tau(\cdot)$ to an HDR image in the training objective. 
Given an HDR image $H$, the $\tau(\cdot)$ is defined as follows:
\begin{equation}
\tau(H) = \frac{\text{log}(1+\mu H)}{\text{log}(1+\mu)},
\label{tonemap}
\end{equation}
where $\mu=5000$ is a parameter of the tone-mapping function. The final loss for training our VQGAN with the OLC is a weighted sum of all losses:
\begin{equation}
\mathcal{L}_{OLC}=\lambda_{rec}\mathcal{L}_{rec}+\lambda_{per}\mathcal{L}_{per}+\lambda_{vq}\mathcal{L}_{vq}+\lambda_{adv}\mathcal{L}_{adv},
\end{equation}
where $\mathcal{L}_{adv}=-\mathbb{E}_{\hat{X}}[D(\hat{X}))]$ is the adversarial loss from discriminator $D$. With the above codebook structure and learning method, OLC is capable of learning the HDR representations over the LDR subspace. 

\begin{figure*}[t]
\centering
\centering
\begin{tikzpicture}
    \node(img) {\includegraphics[width=12.0cm]{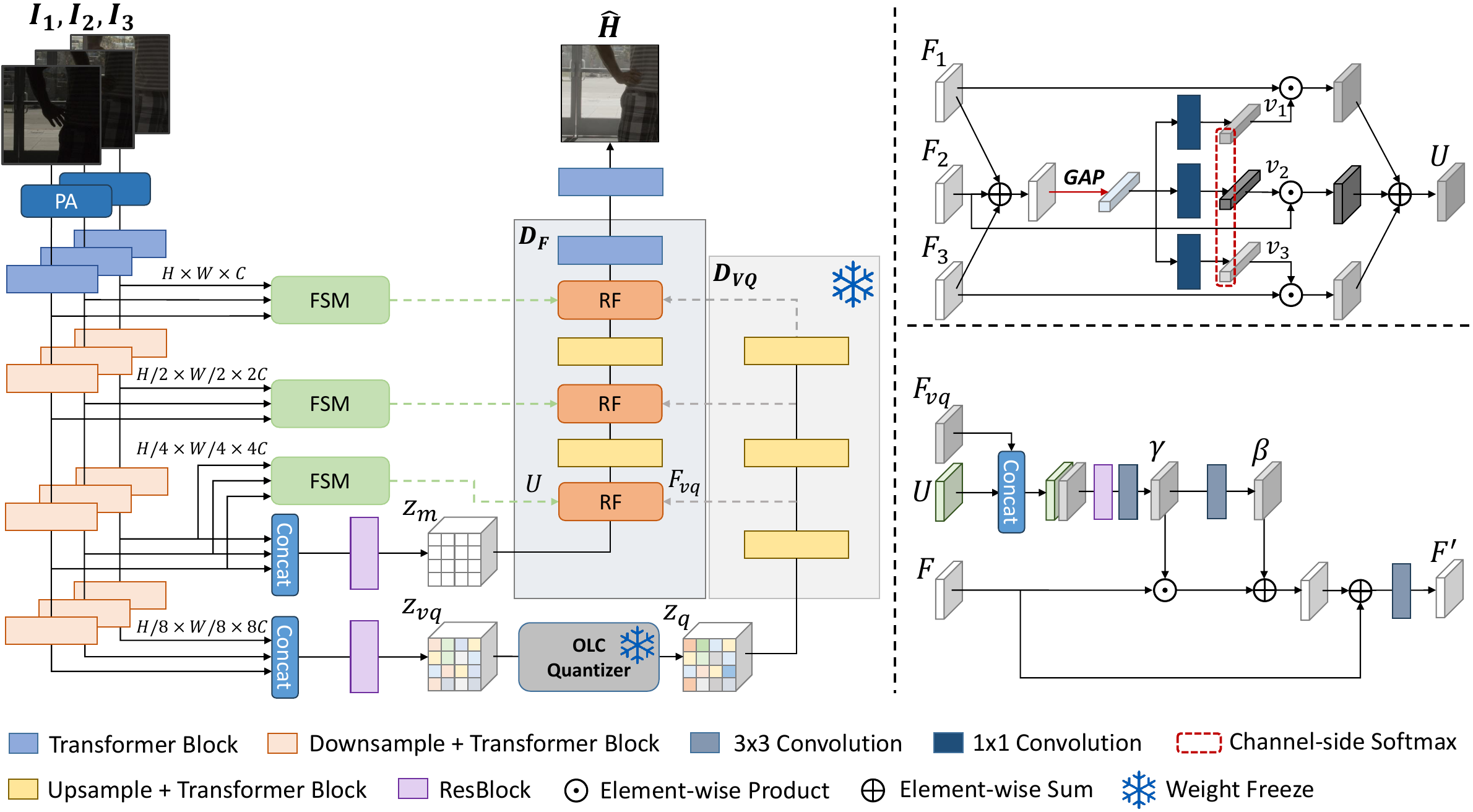}};
    \node[font=\scriptsize](notation_a) at (3.8,3.1) { (a) Frame-Selective Merging (FSM)};
    \node[font=\scriptsize](notation_b) at (3.8,0.3) { (b) Residual Fusing (RF)};
\end{tikzpicture}

\vspace{-0.5em}

\caption{Illustration of proposed dual-decoder HDR network with fidelity decoder $D_{F}$ and pre-trained VQ decoder $D_{VQ}$. The HDR network consists of (a) a Frame-Selective Merging (FSM) unit and (b) a Residual Fusing (RF) unit.} 
\label{hdr_architecture}
\vspace{-1.0em}
\end{figure*}

\subsection{HDR imaging with learned representation}
Following the acquisition of HDR representation through OLC, we introduce an HDR network designed to generate HDR images from multiple LDR images. Specifically, we utilize the acquired HDR representations to enhance the realism of HDR images. To achieve this, we employ a pre-trained codebook and VQ decoder, which is introduced in Sec. 3.1. The learned HDR representation proves beneficial in the HDR reconstruction process by compensating for saturated regions and recovering fine details. However, GAN-based methods often encounter fidelity distortions despite improving perceptual quality which is crucial in multi-exposure HDR imaging. Hence, we propose a network with a dual-decoder structure to address both saturated regions and missing details while preserving image fidelity. 
Given a set of LDR images $L_{i}\in\mathbb{R}^{H\times W\times 3}, i=1,2,3$, we follow previous works that also use corresponding HDR-mapped images as input $I_{i}\in\mathbb{R}^{H\times W\times 6}$ for the network:
\begin{equation}
I_{i}=\left[L_{i}, L_{i}^{\gamma}/t_{i}\right],i=1,2,3,
\end{equation}
where $\gamma=2.2$ is the parameter of the gamma-correction function and $t_{i}$ is the exposure bias (time) of the corresponding LDR frame. We apply a convolution layer to all frames to map them into feature space as: $F_{i} = \text{Conv}(I_{i}), i=1,2,3$. Since input LDR frames are not aligned, we construct the parallel alignment (PA) unit at the initial layer in the HDR network for feature-level alignment.

\begin{figure}[t]
\centering
   \includegraphics[width=12.0cm]{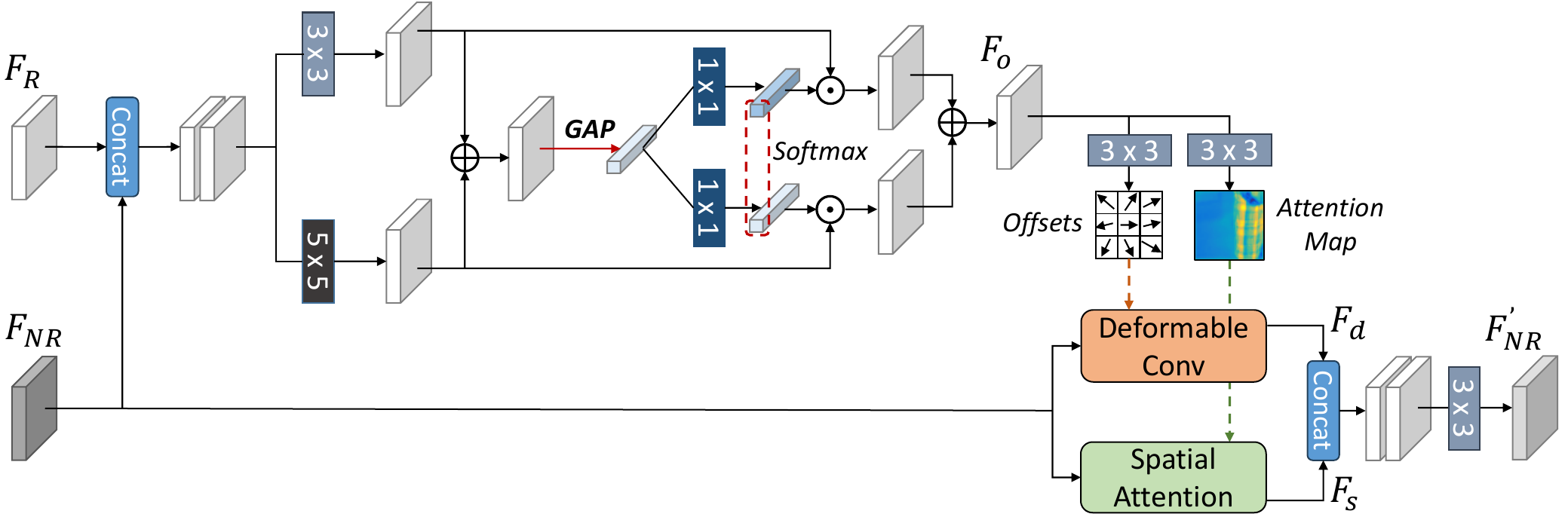}
\caption{Illustration of the Parallel Alignment (PA) unit.} 
\label{parallel_alignment}
\vspace{-1.0em}
\end{figure}

\noindent\textbf{Parallel Alignment.} 
As shown in Fig. \ref{parallel_alignment}, the PA module aligns non-reference frames ($I_{1},I_{3})$ to the reference frame $I_{2}$ in the feature space. Features of both frames are concatenated and processed through an offset module with feature-selective mechanisms and multiple receptive fields. Specifically, $3\times 3$ and $5\times 5$ convolutions are applied to generate an offset feature $F_{o}$, enabling the PA to handle diverse motion differences. Using the offset feature, the PA aligns the non-reference frame feature $F_{NR}$ with deformable convolution and spatial attention. The aligned input features $F_{d}$ and $F_{s}$ are then concatenated to produce the final aligned output $F_{NR}^{\prime}$. This parallel approach with dual alignment methods ensures more accurate alignment. This can be defined as:

\begin{equation}
\begin{aligned}
&F_{d} = DF(F_{NR}, \text{Conv}(F_{o})),\\
&F_{s} = SA(F_{NR}, \text{Conv}(F_{o})),\\
&F_{NR}^{\prime} = \text{Conv}([F_{d},F_{s}]),
\end{aligned}
\end{equation}
where $DF(\cdot)$ and $SA(\cdot)$ denote deformable convolution and spatial attention operation, respectively. Note that we have two non-reference frames $I_{1}, I_{3}$, we define two PA for each non-reference frame, $F_{i}^{\prime} = PA_{i}(F_{i}, F_{2}), i=1,3$. And a convolutional layer applied to reference frame $F_{2}$ as: $F_{2}^{\prime}=\text{Conv}(F_{2})$.

\begin{table*}[t]
\caption{Quantitative comparison on Kalantari \textit{et al.}\cite{kalantari2017deep} and Hu \textit{et al.}\cite{hu2020sensor} dataset. The boldface and underlined numbers denote the best and second-best performances. H.V-2 is HDR-VDP-2 metric. $^{\dagger}$ indicates that the method is excluded from several metrics and experiments since its implementation is not available.} 
\vspace{-1.0em}
{\scriptsize
\begin{center}
\begin{tabular}{c|c|ccccccc}
\toprule
Dataset&Method & PSNR-$\mu$ & PSNR-$\ell$ & PSNR-PU & SSIM-$\mu$ & SSIM-$\ell$ & SSIM-PU & H.V-2\\
\midrule
\multirow{13}{*}{Kalantari~\cite{kalantari2017deep}} & Sen~\cite{sen2012robust}  & 40.95 & 38.30 & 34.44 & 0.9829 & 0.9745 & 0.9783 & 59.38\\
&Kalantari~\cite{kalantari2017deep} & 42.74 & 41.23 & 36.35 & 0.9888 & 0.9846 & 0.9843 & 64.42 \\ 
&DeepHDR~\cite{wu2018deep}& 41.91 & 40.36 & 35.52 & 0.9770 & 0.9602 & 0.9805 & 64.78 \\ 
&AHDRNet~\cite{yan2019attention}& 43.70 & 41.17 & 37.37 & 0.9904 & 0.9856 & 0.9869 & 65.11 \\ 
&NHDRRNet$^{\dagger}$~\cite{yan2020deep}& 42.41 & - & - & 0.9887 & - & - & - \\ 
&HDR-GAN~\cite{niu2021hdr}& 43.92 & 41.57 & 37.47 & 0.9905 & 0.9865 & 0.9870 & 65.58 \\
&ADNet~\cite{liu2021adnet}& 43.97 & 41.78 & 37.62 & 0.9905 & 0.9882 & 0.9867 & 65.84 \\
&TransHDR$^{\dagger}$~\cite{song2022selective} & 44.10 & 41.70 & - & 0.9909 & 0.9872 & - & - \\
&CA-ViT~\cite{liu2022ghost} & 44.32 & 42.18 & 37.73 & 0.9916 & 0.9884 & 0.9878 & 66.33 \\
&HFT$^{\dagger}$~\cite{chen2023improving} & 44.45 & 42.14 & - & 0.9920 & 0.9880 & - & 66.32 \\
&SCTNet~\cite{tel2023alignment} & 44.47 & 42.33 & \underline{37.95} & \underline{0.9922} & 0.9885 & \underline{0.9887} & \underline{66.40} \\

&HyHDRNet$^{\dagger}$~\cite{yan2023unified} & \underline{44.64} & \underline{42.47} & - & 0.9915 & \underline{0.9894} & - & 66.03 \\

&\textbf{Proposed} & \textbf{44.89} & \textbf{42.60} & \textbf{38.32} & \textbf{0.9935} & \textbf{0.9898} & \textbf{0.9899} & \textbf{66.69} \\
\midrule

\multirow{10}{*}{Hu~\cite{hu2020sensor}}&Sen~\cite{sen2012robust} & 31.51 & 33.45 & 30.81 & 0.9533 & 0.9630 & 0.9783 & 59.38\\
&Kalantari~\cite{kalantari2017deep} & 42.74 & 41.23 & 36.35 & 0.9888 & 0.9846 & 0.9843 & 63.72 \\ 
&DeepHDR~\cite{wu2018deep} & 41.88 & 41.96 & 35.81 & 0.9790 & 0.9856 & 0.9860 & 63.15 \\
&AHDRNet~\cite{yan2019attention} & 46.87 & 50.70 & 41.26 & 0.9959 & 0.9983 & 0.9956 & 64.29 \\
&HDR-GAN~\cite{niu2021hdr} & 46.69 & 50.42 & 41.02 & 0.9958 & 0.9988 & 0.9954 & 64.33\\
&ADNet~\cite{liu2021adnet}& 47.27 & 51.83 & 41.44 & 0.9961 & 0.9988 & 0.9957 & 64.47 \\
&CA-ViT~\cite{liu2022ghost} & 47.98 & 52.12 & 41.68 & \underline{0.9967} & 0.9990 &0.9960& 64.67 \\
&SCTNet~\cite{tel2023alignment} & 48.18 & \underline{52.15} & \underline{41.72} & \underline{0.9967} & \underline{0.9991} &\underline{0.9962} & \underline{64.84} \\
&HyHDRNet$^{\dagger}$~\cite{yan2023unified} & \underline{48.46} & 51.91 & - & 0.9959 & \underline{0.9991} & - & -\\
&\textbf{Proposed} & \textbf{48.73} & \textbf{52.39} & \textbf{42.47} & \textbf{0.9970} & \textbf{0.9992} & \textbf{0.9966} & \textbf{65.12}\\
\bottomrule

\end{tabular}
\end{center}
}
\label{comparison}
\vspace{-2.0em}
\end{table*}

Following the alignment of non-reference frame features, we establish individual multi-scale encoders to extract features from each LDR frame. Each encoder processes the frame feature $F_{i}^{\prime}\in\mathbb{R}^{H\times W\times C}$ and progressively reduces the spatial size to $\frac{H}{8}\times \frac{W}{8}\times 8C$. As depicted in Fig. ~\ref{hdr_architecture}, we combine frame features at both $\frac{H}{4}\times \frac{W}{4}$ and $\frac{H}{8}\times \frac{W}{8}$ scales for the fidelity decoder $D_{F}$ and pre-trained VQ decoder $D_{VQ}$, respectively. Given that the pre-trained VQ decoder is trained on $\frac{H}{8}\times \frac{W}{8}$ spatial size, we input the same spatial size of the quantized merged feature $z_{q}=\mathcal{Q}(z_{vq}, \mathcal{Z})\in\mathbb{R}^{\frac{H}{8}\times \frac{W}{8}\times 8C}$ into the VQ decoder to minimize discrepancies. Note that we use full codebook $\mathcal{Z}$ to quantize since we target reconstructing HDR images in the HDR network. Conversely, for the fidelity decoder, we input merged features $z_{m}\in\mathbb{R}^{\frac{H}{4}\times \frac{W}{4}\times 4C}$ with a less reduced scale to preserve structural information. Specifically, the fidelity decoder incorporates features from the VQ decoder and a frame context feature from the encoding stage. Different from existing methods that solely deliver the reference frame feature with a skip connection, we introduce a Frame-Selective Merging (FSM) unit, which aggregates encoded frame contexts for delivering richer frame information to the decoder.



\noindent\textbf{Frame-Selective Merging.} In Fig.~\ref{hdr_architecture}(a), we illustrate the Frame-Selective Merging (FSM) unit. Inspired by ~\cite{li2019selective}, FSM employs attention-based mechanisms to aggregate frame features $F_{i}$. It first combines input features through summation, then applies global average pooling and a $1\times 1$ convolution to generate a feature vector $v$. This vector undergoes three individual $1\times 1$ convolutions and channel-wise softmax to produce attention vectors $v_{i}$ for each frame. The attention vectors $v_{i}$ are then multiplied by their corresponding frame features, and the processed features are summed to produce the merged context $U=\sum_{i}(F_{i}\odot v_{i})$. By selecting valid features from each frame, FSM effectively merges frame context, thereby supporting the decoding process.

\begin{figure*}[t]
\centering
\includegraphics[width=12cm]{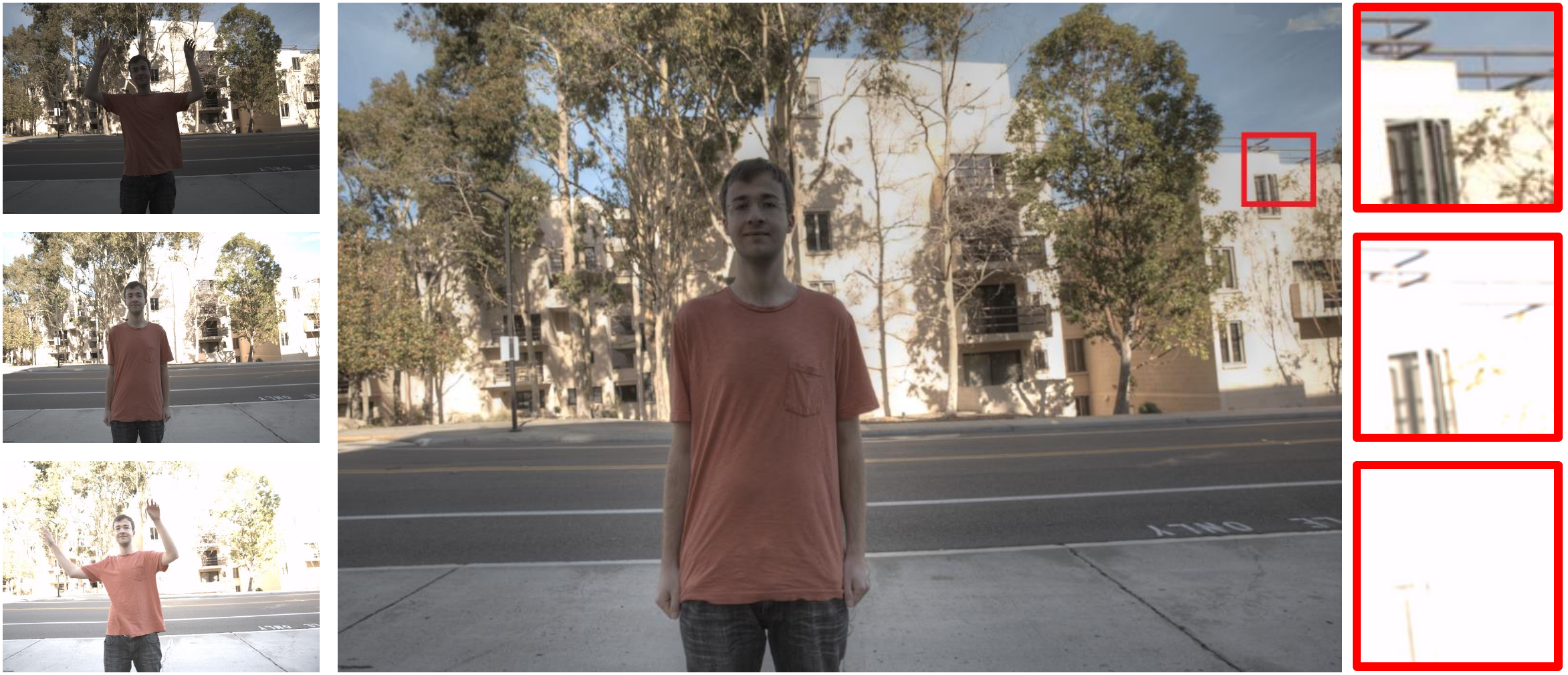}

\vspace{-0.3em}

\raggedright 
\hspace{3em}LDRs \hspace{7em}Our tone-mapped HDR image \hspace{6.5em}Patches

\vspace{0.0em}

\centering
\includegraphics[width=12cm]{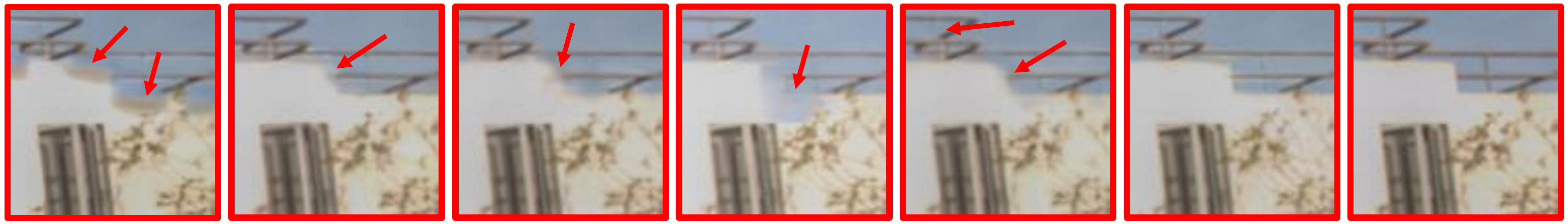}

\vspace{-0.2em}

\raggedright
\hspace{1.5em}Yan~\cite{yan2019attention}\hspace{2.2em}Niu~\cite{niu2021hdr}\hspace{2em}Liu~\cite{liu2021adnet}\hspace{2em}Liu~\cite{liu2022ghost}\hspace{2.0em}Tel~\cite{tel2023alignment}\hspace{2.3em}Ours\hspace{3.5em}GT

\vspace{-0.5em}
\caption{Visual comparison on a test sample in Kalantari's \cite{kalantari2017deep} dataset.}
\vspace{-1.0em}
\label{visual_comparison_kalantari_1}
\end{figure*}

\noindent\textbf{Residual Fusing.} As we stated earlier, our HDR network features a dual-decoder structure. We use a pre-trained VQ decoder $D_{VQ}$ with OLC and add a fidelity decoder $D_{F}$ for HDR reconstruction. To leverage the VQ decoder's HDR representation capabilities, we propose a Residual Fusing (RF) module. As shown in Fig.~\ref{hdr_architecture}(b), RF takes intermediate features $F_{vq}$ from $D_{VQ}$ and merged contexts $U$ from FSM to fuse internal features in $D_{F}$. Both $F_{vq}$ and $U$ are concatenated and fed into a resblock to produce parameter features $\gamma$ and $\beta$. RF then fuses the input feature with $\gamma$ and $\beta$ through affine transformation, finally producing output feature $F^{\prime}$ with a residual connection. This can be defined as:
\begin{equation}
\begin{aligned}
&\gamma,\beta = \text{Conv}([U,F_{vq}]),\\
&F^{\prime}=(\gamma\odot F + \beta) + F.
\end{aligned}
\end{equation}

With this residual fusing method, RF is able to incorporate VQ features and context while retaining image fidelity with the residual connection.

The training objective of our HDR network is the combination of three losses: 1) reconstruction loss $\mathcal{L}_{rec}$ for maintaining data fidelity; 2) perceptual loss $\mathcal{L}_{per}$ for producing realistic details; 3) mapping loss $\mathcal{L}_{map}$ for mapping extracted features to code vectors in the learned codebook. Given the ground-truth HDR image $H$ and a predicted HDR image $\hat{H}$, the $\mathcal{L}_{rec}, \mathcal{L}_{per}$ can be defined as:
\begin{equation}
\mathcal{L}_{rec}=\|\tau(H)-\tau(\hat{H})\|_{1}, \mathcal{L}_{per}=\|\phi(\tau(H))-\phi(\tau(\hat{H}))\|_{1},
\end{equation}
where $\phi(\cdot)$ is pre-trained VGG-16 network \cite{simonyan2014very}. The mapping loss $\mathcal{L}_{map}$ calculates the distance between the extracted feature $z_{gt}\in\mathbb{R}^{\frac{H}{8}\times\frac{W}{8}\times8C}$ in the HDR network and ground-truth VQ representation $z_{gt}=\mathcal{Q}(E(H),\mathcal{Z})$, defined as:
\begin{equation}
    \mathcal{L}_{map}=\|z_{vq}-z_{gt}\|_{2}^{2}.
\end{equation}
The final loss $\mathcal{L}_{HDR}$ is weighted sum of all losses :
\begin{equation}
\mathcal{L}_{HDR}=\mathcal{L}_{rec}+\lambda_{per}\mathcal{L}_{per}+\lambda_{map}\mathcal{L}_{map}.
\end{equation}
\begin{figure*}[t]
\centering
    \begin{minipage}{.485\textwidth}
            \begin{center}
                \includegraphics[width=\textwidth]{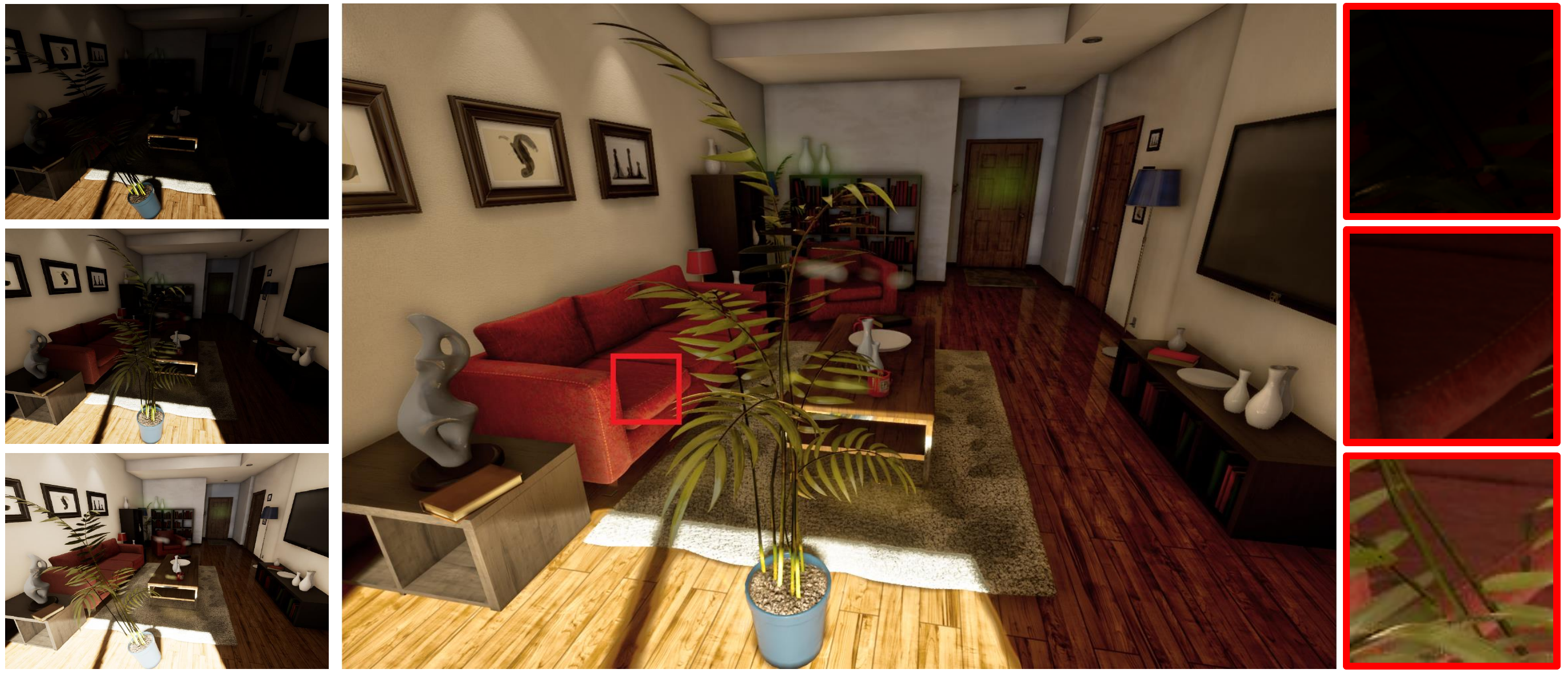}
                
                \raggedright\hspace{0.6em}LDRs \hspace{1.2em} Result HDR image \hspace{0.5em} Patches

                \vspace{0.1em}
                
                \centering
                \includegraphics[width=\textwidth]{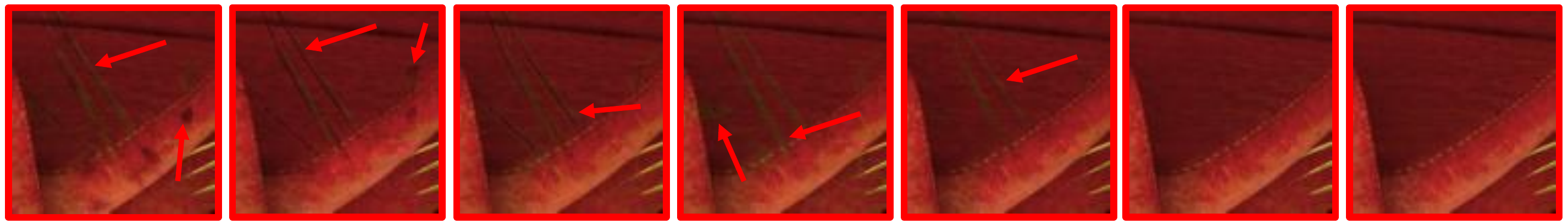}

                \scriptsize\raggedright\hspace{0.0em}Yan\cite{yan2019attention}\hspace{0.2em}Niu\cite{niu2021hdr}\hspace{0.2em}Liu\cite{liu2021adnet}\hspace{0.2em}Liu\cite{liu2022ghost}\hspace{0.2em}Tel\cite{tel2023alignment}\hspace{0.4em}Ours\hspace{1.0em}GT

                \vspace{0.1em}
                \normalsize
                \footnotesize\centering (a)  Test sample in Hu's dataset.
            \end{center}
        \end{minipage}
        \hfill
        \begin{minipage}{.50\textwidth}
            \begin{center}
                \includegraphics[width=\textwidth]{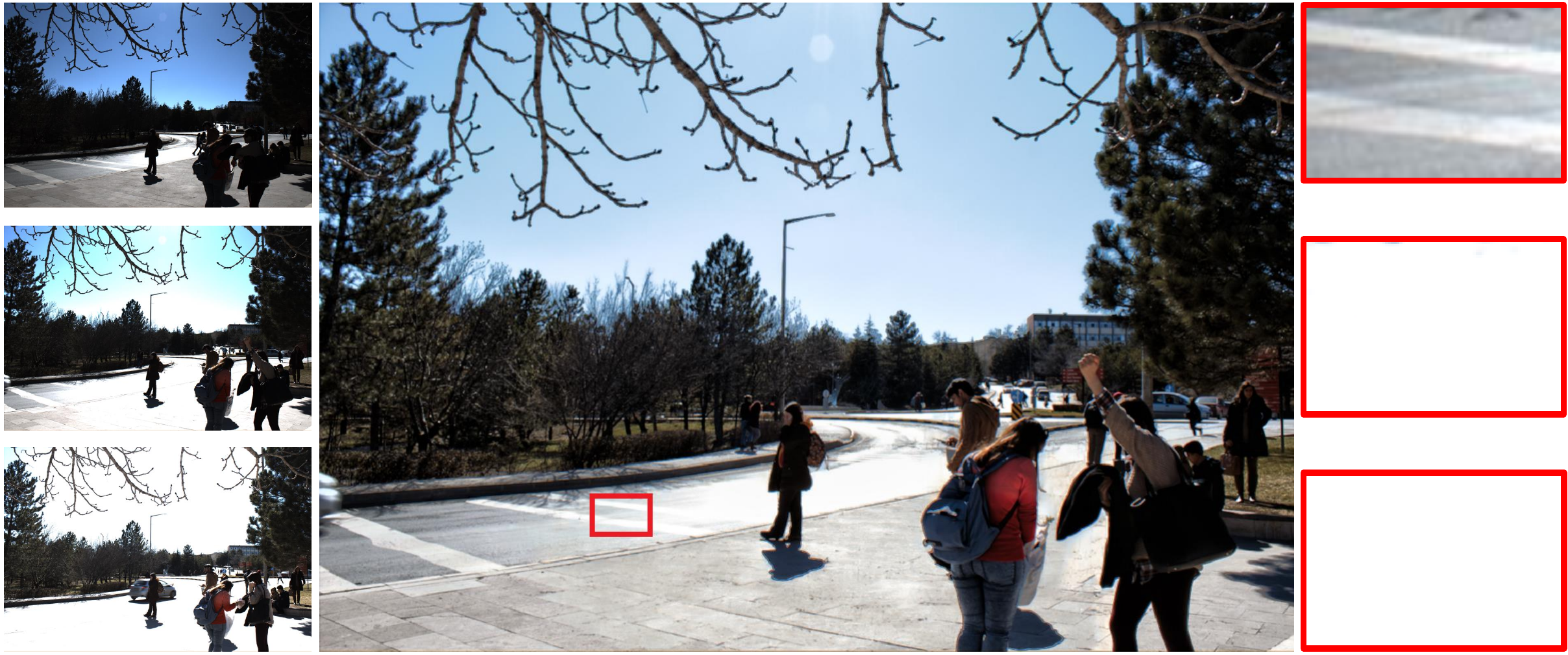}
                
                \raggedright\hspace{0.6em}LDRs \hspace{1.6em} Result HDR image \hspace{0.7em} Patches

                \vspace{0.1em}
                
                \centering
                \includegraphics[width=\textwidth]{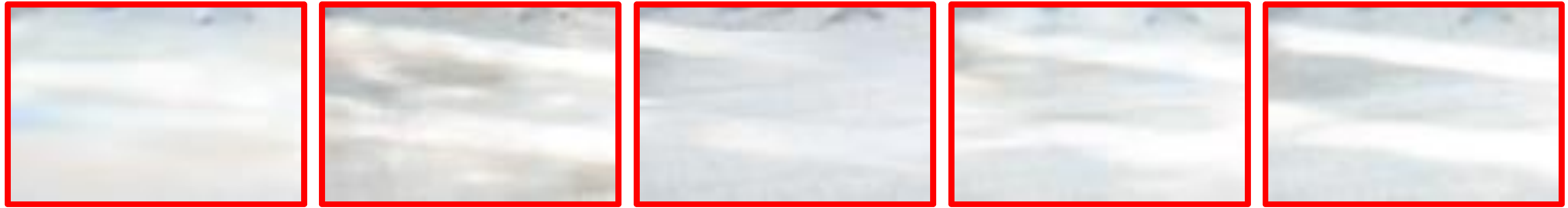}
                
                \scriptsize\raggedright
                \hspace{0.8em}Yan\cite{yan2019attention}\hspace{1.5em}Niu\cite{niu2021hdr}\hspace{1.5em}Liu\cite{liu2022ghost}\hspace{1.5em}Tel\cite{tel2023alignment}\hspace{1.7em}Ours

                \vspace{0.1em}
                \footnotesize\centering(b) Test sample in Tursun's dataset.
            \end{center}
        \end{minipage}
        \caption{Visual comparison on test samples in (a) Hu's dataset and (b) Tursun's dataset. Note that samples in Tursun's dataset has no ground-truth HDR images.}
        \label{visual_comparison_samsung}
\vspace{-1.0em}
\end{figure*}

\section{Experiments}
\label{sec:experiments}
\vspace{-0.1em}
\subsection{Dataset and Metrics} 
\noindent\textbf{Dataset.} We train and test our method on Kalantari \textit{et al.}'s dataset \cite{kalantari2017deep} and Hu \textit{et al.}'s dataset~\cite{hu2020sensor}. Specifically, Kalantari \textit{et al.}'s dataset consists of 74 samples for training and 15 samples for testing. Each data pair contains three LDR images that are captured with \{-2, 0, +2\} or \{-3, 0, +3\} of exposure bias sets and a single HDR image. 
Hu \textit{et al.}'s dataset~\cite{hu2020sensor} synthesized with the game engine, and captured with an exposure bias of \{$-2,0,+2$\}. 

\noindent\textbf{Evaluation Metrics.} We compute metrics on both results linear HDR image $\hat{H}$ and tone-mapped HDR image $\tau(\hat{H})$. The PSNR-$\ell$, SSIM-$\ell$ are calculated between linear HDR image $H$, $\hat{H}$ and PSNR-$\mu$, SSIM-$\mu$ are calculated between tone-mapped images $\tau(H), \tau(\hat{H})$. Furthermore, we also measure HDR-VDP-2~\cite{mantiuk2011hdr}, which evaluates the quantitative quality of HDR images on specified display and luminance conditions. Lastly, we report the PU21~\cite{azimi2021pu21} metric, which measure the similarity between perceptually uniform values of the HDR images.

\vspace{-0.1em}

\subsection{Training Details} For training both the OLC and the HDR network, we crop patches of size $256\times 256$ with a stride of 64 from training samples. Further, we also apply a set of augmentation, including horizon/vertical flipping and rotation. All experiments are implemented with the Pytorch framework and a single NVIDIA RTX 3090 Ti GPU. We adopt Adam optimizer \cite{kingma2014adam} with 1e-4 learning rate for training generators in OLC and HDR network. For the discriminator in VQGAN, a learning rate of 4e-4 is set. The number of code vectors in the OLC is set as $K=1024$ and the base channel size of the HDR network is $C=32$.

\subsection{Comparison with Previous Methods}
\vspace{-0.2em}
\textbf{Quantitative Comparison.}
Tab.~\ref{comparison} shows a quantitative comparison with previous methods on Kalantari's dataset\cite{kalantari2017deep} and Hu's dataset~\cite{hu2020sensor}. Generally, deep learning-based methods~\cite{kalantari2017deep,wu2018deep,yan2019attention,liu2021adnet,niu2021hdr} show improved performance compared to patch-based \cite{hu2013hdr,sen2012robust} algorithms. Furthermore, Transformer-based methods~\cite{liu2022ghost,yan2023unified,chen2023improving,tel2023alignment} outperform previous methods by notable margins. 
Our method achieves the best performance on most metrics, including HDR-VDP-2 and PU21. This result implies our method is not only producing more realistic HDR images but also robust on certain display and luminance conditions. 

\noindent\textbf{Qualitative Comparison.} We further evaluate the qualitative results in Fig.~\ref{visual_comparison_kalantari_1} and Fig.~\ref{visual_comparison_samsung}. Note that we use a tone-mapping function of Photomatix to visualize HDR images. Fig. \ref{visual_comparison_kalantari_1} displays the ability to reconstruct heavily saturated regions. AHDRNet~\cite{yan2019attention}, ADNet~\cite{liu2021adnet}, and HDR-GAN~\cite{niu2021hdr} produce blurry detail component and edges regions. CA-ViT~\cite{liu2022ghost} and SCTNet~\cite{tel2023alignment} show the resulting image with better-detailed regions, but there are distorted region remains on the edges. In contrast, our method produces clear edges and fine details without distortion. In Fig. \ref{visual_comparison_samsung} (a), a large motion difference exists between LDR frames. Different from other methods that leave ghosting artifacts on moving objects, our method effectively address misalignment with PA modules and produces result HDR images without undesired artifacts. We also compare our method on the Tursun \textit{et al.}~\cite{tursun2016objective} dataset, which has no ground-truth HDR image in Fig. \ref{visual_comparison_samsung} (b). Since the scene information in the reference frame and high exposure frame was severely lost due to over-exposure, other methods failed to compensate for saturated regions from valid regions in other frames. In contrast, our method shows more realistic HDR images in extreme cases.  We report additional quantitative and qualitative results in the supplementary materials.

\begin{figure}[t]

\begin{tikzpicture}
    \node(img1) at (-2.5,0) {\includegraphics[width=5.9cm]{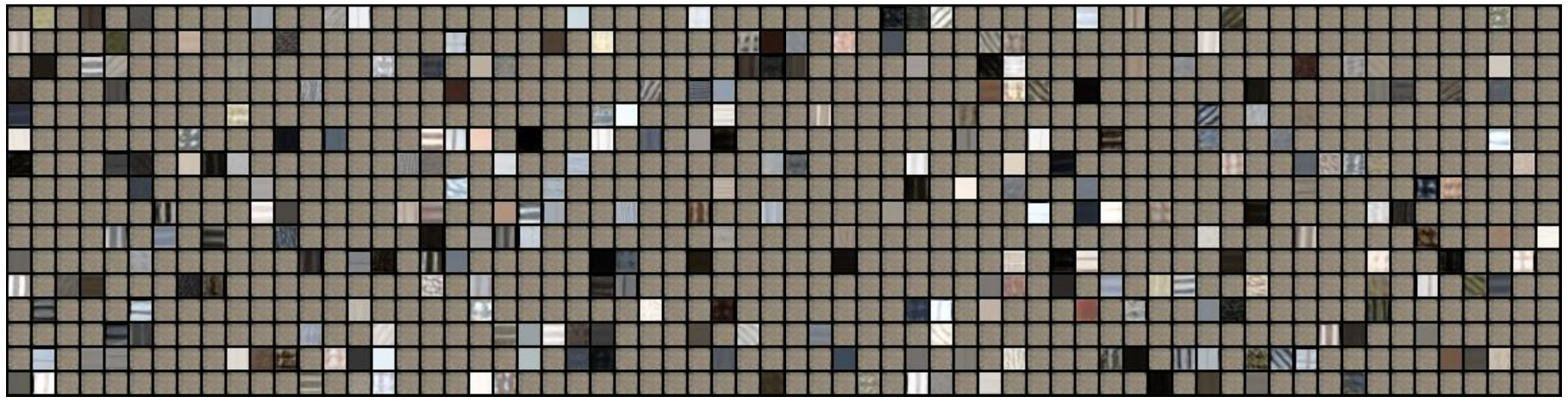}};
    \node(img2) at (3.7,0) {\includegraphics[width=5.9cm]{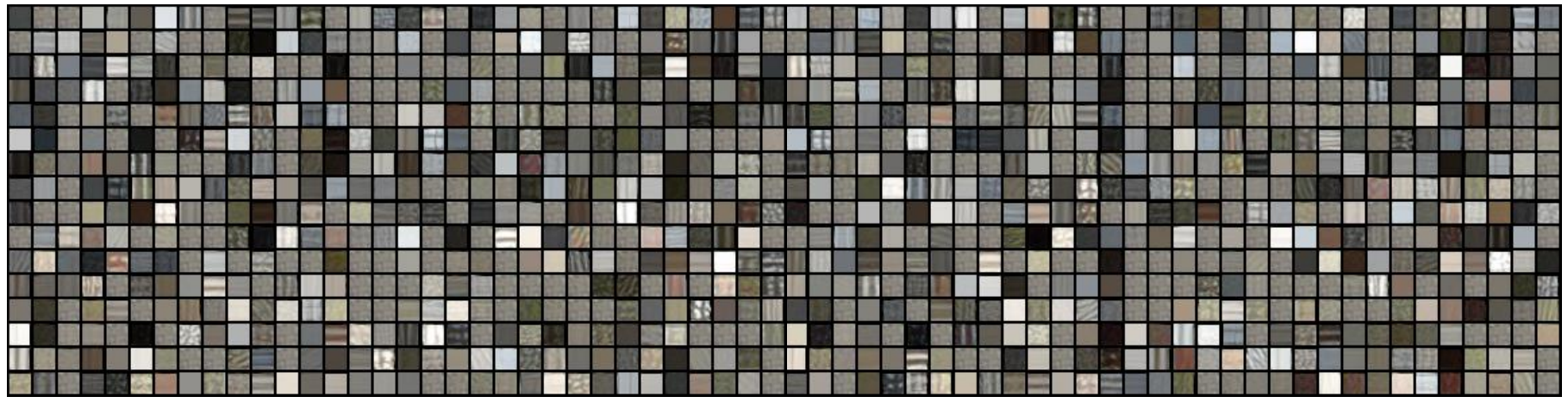}};
    
    \node(img3) at (-2.45,-1.8) {\includegraphics[width=5.7cm]{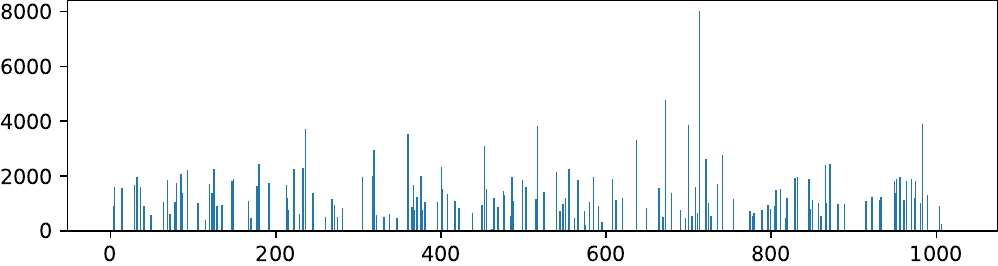}};
    \node(img4) at (3.75,-1.8) {\includegraphics[width=5.7cm]{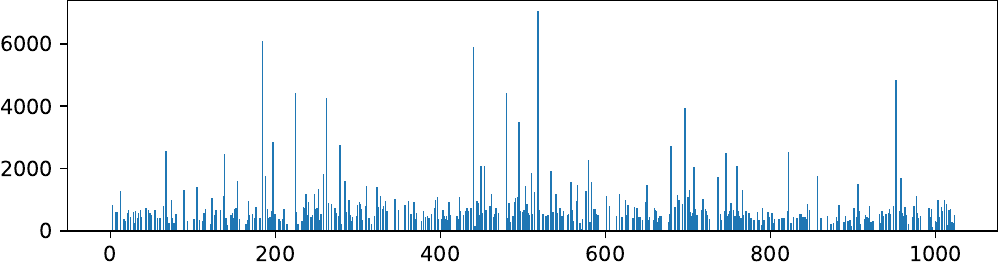}};
    \node(notation_a) at (-2.5,-2.7) {\tiny Code index};
    \node(notation_b) at (3.7,-2.7) {\tiny Code index};
    \node[rotate=90](notation_c) at (-5.4,-1.7) {\tiny Count};
    \node[rotate=90](notation_d) at (0.8,-1.7) {\tiny Count};

    \node(notation_e) at (-2.5,-3.1) {\scriptsize (a) Vanilla VQ Codebook};
    \node(notation_f) at (3.8,-3.1) {\scriptsize (b) Overlapped Codebook};
\end{tikzpicture}
\vspace{-2em}
\caption{Code vector visualization (first row) and distribution (second row) in the vanilla VQ codebook and proposed Overlapped codebook (OLC). }
\label{fig:codebooks}
\vspace{-1em}
\end{figure}
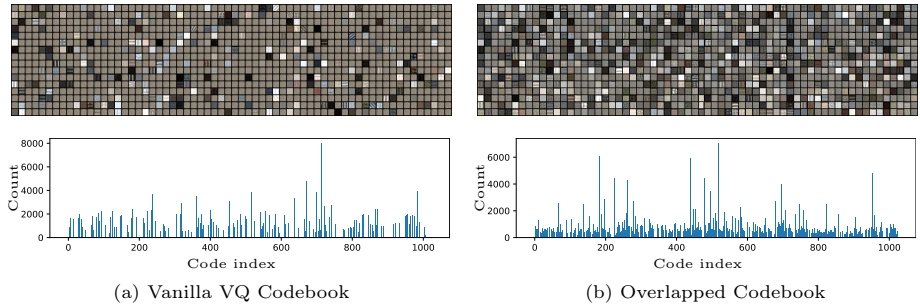

\begin{table}[t]
\centering
    \begin{minipage}{0.52\textwidth}
    \footnotesize
    \centering
    \vspace{-1.5em}
    \caption{Performance on Test samples in~\cite{kalantari2017deep} with vanilla codebook and OLC. $K$ denotes the number of code vectors.}
        \begin{tabular}{c|ccc}
        \toprule
        Method &  PSNR-$\mu$ & PSNR-$\ell$ & H.V-2\\
        \midrule
         Vanilla (K=512)& 44.38 & 42.20 & 66.31  \\
         OLC (K=512)& 44.55 & 42.36 & 66.42  \\
         Vanilla (K=1024)& 44.57 & 42.32 & 66.44 \\
         OLC (K=1024)& 44.89 & 42.60 & 66.69  \\
         \bottomrule
    \end{tabular}
    \label{tab:olc_ablation}
    \end{minipage}\hfill
    \begin{minipage}{0.44\textwidth}
    \centering
           
    \begin{tikzpicture}
    \node(img1) at (0,0) {\includegraphics[width=4.2cm]{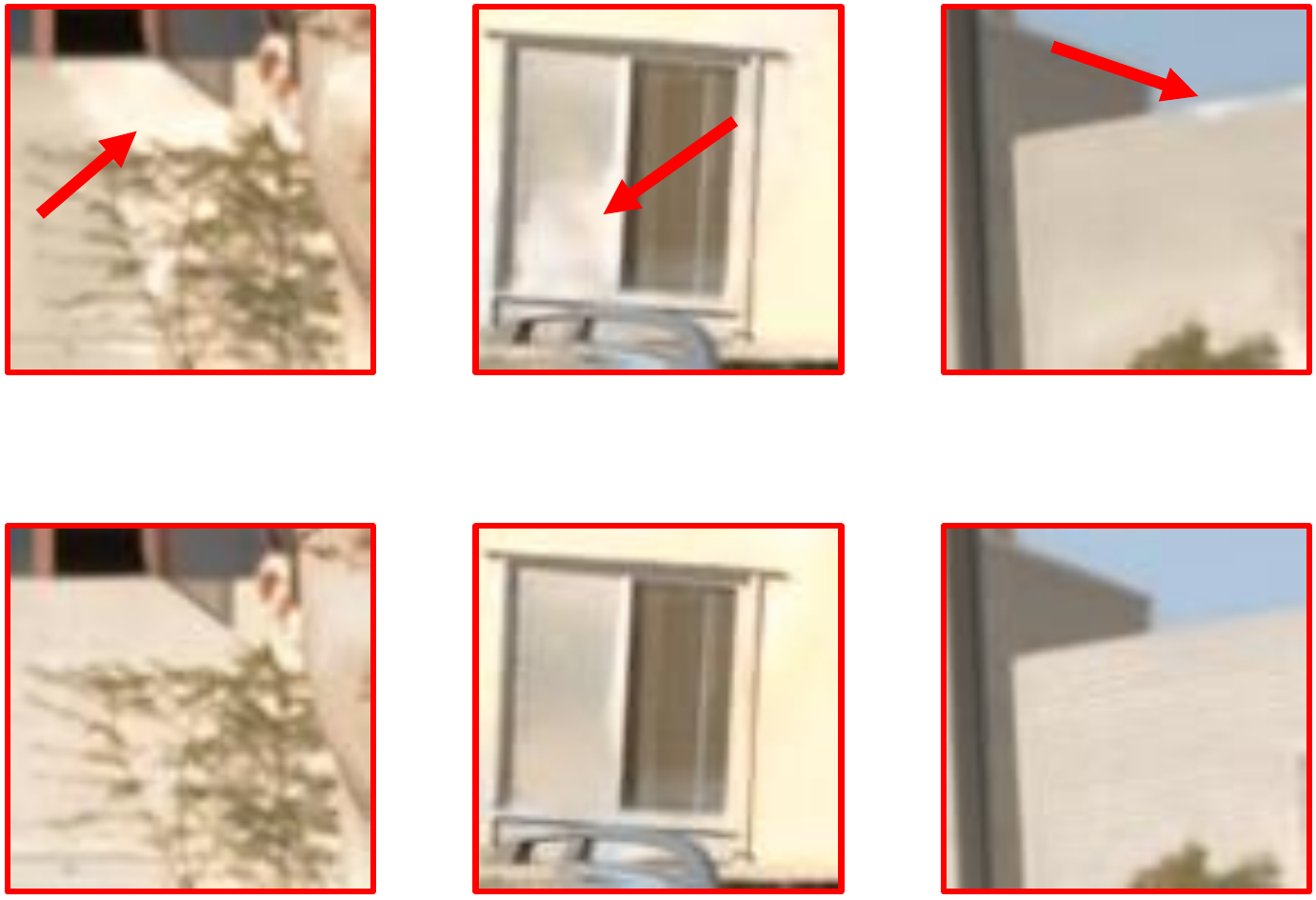}};
    \node(notation_a) at (0,0) {\scriptsize w/ Vanilla codebook};
    \node(notation_a) at (0,-1.60) {\scriptsize w/ Overlapped codebook (OLC)};
    \end{tikzpicture}
        
        \captionof{figure}{Visual comparison on vanilla codebook and OLC (K=1024).}
    \label{fig:olc_ablation}
    \end{minipage}
\vspace{-2.0em}
\end{table}

\subsection{Analysis on the proposed OLC}

As previously discussed, proposed OLC significantly enhances the capacity to learn implicit HDR representations. In Fig.~\ref{fig:codebooks}, we provide visualizations of code vectors within the pre-trained VQGAN framework and display the code index distribution for reconstructing HDR images. It's important to note that both the vanilla VQ codebook (a) and the OLC (b) are trained under identical conditions, including training iteration and network settings. The visualization illustrates that the proposed OLC explores a more diverse range of HDR representations, learning additional valid code vectors and utilizing them to restore HDR images. Furthermore, we compare the performance of OLC with the vanilla codebook in Tab.~\ref{tab:olc_ablation}. OLC demonstrates superior performance in reconstructing HDR images, particularly with a larger codebook size ($K$). In Fig. ~\ref{fig:olc_ablation}, we showcase predicted HDR patches with the vanilla codebook (first row) and OLC (second row). Compared to the vanilla codebook, OLC exhibits enhanced capability in restoring saturated and detailed regions. These results affirm that our OLC offers improved representation learning ability, consequently enhancing performance without additional computational burden in reconstructing HDR images.

\begin{table}[t]
\vspace{-1.0em}
\centering

\begin{minipage}{.65\textwidth}
    \caption{Ablation on proposed components. \textit{Sum} and \textit{Concat} in variants 3, 4 denote the frame merging method.}  
\centering
\footnotesize      
\begin{tabular}{l|ccc}

\toprule

Method & PSNR-$\mu$ & PSNR-$\ell$ & H.V-2\\
\midrule
1. Baseline & 43.92 & 41.77 & 65.79\\
2. + PA & 44.20 & 41.94 & 66.02\\
3. + PA + \textit{Sum} & 44.31 &  42.11 & 66.15\\
4. + PA + \textit{Concat} & 44.38 &  42.22 & 66.22\\
5. + PA + FMU & 44.49 &  42.30 & 66.35\\
6. + PA + FMU + $D_{VQ}$ & 44.74 & 42.51 & 66.60\\
7. + PA + FMU + $D_{VQ}$ + RF & 44.89 & 42.60 & 66.69\\
\bottomrule
\end{tabular}
\label{tab:ablation}
    \end{minipage}\hfill
    \begin{minipage}{.32\textwidth}
    \centering
    \vspace{1.0em}
    \begin{tikzpicture}
    \node(img1) at (0,-0) {\includegraphics[width=3.7cm]{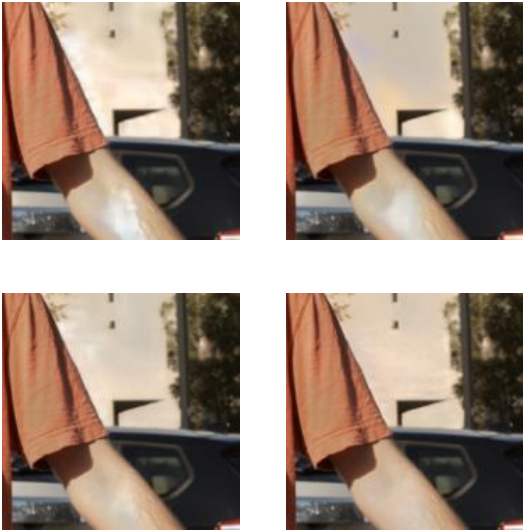}};
    \node(notation_a) at (-1,0.05) {\scriptsize Baseline};
    \node(notation_a) at (1,0.05) {\scriptsize Variant 2};
    \node(notation_a) at (-1,-2) {\scriptsize Variant 5};
    \node(notation_a) at (1,-2.05) {\scriptsize Proposed};
    \end{tikzpicture}
    \captionof{figure}{Visual comparison on variants in ablation.}
    \label{fig:ablation}
    \end{minipage}
    \vspace{-2.0em}
  \end{table}

\subsection{Impact of proposed modules}
In Tab.~\ref{tab:ablation} and Fig.~\ref{fig:ablation}, we conduct an ablation study on Kalantari's dataset to demonstrate the effectiveness of the proposed modules in the HDR network. The Baseline model consists of an encoder and fidelity decoder. Variants 3 and 4 merge frame contexts by summing ($U=F_{1}+F_{2}+F_{3}$) or concatenating ($U=\text{Conv}([F_{1},F_{2},F_{3}])$) instead of using the FMU. Variants 3-6 also incorporate merged context $U$ or VQ feature $F_{vq}$ without the RF module. Variant 2, with PA modules, reduces ghosting artifacts and improves quality in misalignment regions. Variant 5, with FMU modules, better compensates for saturated regions. Variant 7, the proposed network that incorporating all proposed components, achieves the best performance, producing more realistic HDR images. These results validate the effectiveness of each proposed module and the pre-trained VQ component in enhancing HDR reconstruction performance.

\section{Conclusion}
\label{sec:conclusion}
We proposed an Overlapped Codebook (OLC) scheme for multi-exposure HDR imaging, which effectively learns implicit HDR representations within the VQGAN framework by modeling the HDR generation process in exposure bracketing. Additionally, we introduced a dual-decoder HDR network that leverages these acquired HDR representations from the pre-trained OLC to produce high-quality HDR images. Our network includes a parallel alignment module to correct misalignment among LDR frames and features frame-selective merging and residual fusing modules to integrate HDR representations with valid frame contexts during decoding. Extensive experiments demonstrate significant improvements with our method on benchmark datasets.

\vspace{2em}

\noindent\textbf{Acknowledgement.} This work was supported by Institute of Information \& communications Technology Planning \& Evaluation (IITP) grant funded by the Korea government(MSIT) [NO.RS-2021-II211343, Artificial Intelligence Graduate School Program (Seoul National University)] [No.RS-2021-II212068, Artificial Intelligence Innovation Hub], and in part by Samsung Electronics Co., Ltd.

%
%


%
%
%
%

\end{document}